\documentclass[10pt, conference]{IEEEtran}
\IEEEoverridecommandlockouts

\usepackage[utf8x]{inputenc}
\usepackage[T1]{fontenc}
\usepackage[english]{babel}
\usepackage{microtype}
\usepackage{setspace, balance}
\usepackage{amsmath,amssymb,amsfonts, bm}
\usepackage{graphicx, caption, subcaption}
\usepackage{booktabs, threeparttable, multirow}
\usepackage[inline]{enumitem}
\usepackage{hyperref, url, xcolor}
\usepackage{siunitx}
\usepackage[printwatermark]{xwatermark}
\usepackage{listings}
\usepackage[ruled, vlined]{algorithm2e}

\graphicspath{{images/}}
\DeclareGraphicsExtensions{.pdf,.jpeg,.png}

\definecolor{dark-red}{rgb}{.6, .15, .15}
\definecolor{dark-blue}{rgb}{.15, .15, .55}
\definecolor{accent1}{HTML}{24B9FC}
\definecolor{accent2}{HTML}{9ECD67}
\definecolor{accent3}{HTML}{FFA929}
\definecolor{light-bg}{HTML}{B2B2B2}

\newcommand{\ra}[1]{\renewcommand{\arraystretch}{#1}}

\newcommand{\code}[1]{\texttt{#1}}

\hypersetup{
	colorlinks,
  	linkcolor={dark-red},
	citecolor={dark-blue},
  	urlcolor={black}
}

\newif\ifunblind
\unblindtrue

\begin{document}

\title{Towards Semantic Clone Detection \\ via Probabilistic Software Modeling}

\ifunblind
\author{%
	\IEEEauthorblockN{Hannes Thaller, Lukas Linsbauer, Alexander Egyed}
	\IEEEauthorblockA{Institute for Software Systems Engineering\\
		Johannes Kepler University Linz, Austria\\
		\{hannes.thaller, lukas.linsbauer, alexander.egyed\}@jku.at}
}
\else
\author{\IEEEauthorblockN{Author}
	\IEEEauthorblockA{Institute\\
		Affiliation, Country\\
		email@address.com}
}
\fi

\maketitle

\begin{abstract}
Semantic clones are program components with similar behavior, but different textual representation.
Semantic similarity is hard to detect, and semantic clone detection is still an open issue.
We present semantic clone detection via Probabilistic Software Modeling (PSM) as a robust method for detecting semantically equivalent methods.
PSM inspects the structure and runtime behavior of a program and synthesizes a network of Probabilistic Models (PMs).
Each PM in the network represents a method in the program and is capable of generating and evaluating runtime events.
We leverage these capabilities to accurately find semantic clones.
Results show that the approach can detect semantic clones in the complete absence of syntactic similarity with high precision and low error rates.
\end{abstract}

\begin{IEEEkeywords}
clone detection, semantic clone detection, probabilistic modeling, multivariate testing, software modeling, static code analysis, dynamic code analysis, runtime monitoring, inference, simulation, deep learning
\end{IEEEkeywords}

\IEEEpeerreviewmaketitle







\section{Introduction}
Copying and pasting source code fragments leads to code clones.
Code clones are considered an anti-pattern as they increase maintenance costs, promote bad software design, and propagate bugs \cite{Mayrand1996, Monden2002, Martin2009, Fowler1999, Hunt2000, Chou2001, Li2006, Geiger2006}.
Code clones are traditionally split into four categories.
Type~1-3~\cite{Roy2007, Bellon2007, Rattan2013} code clones are textual copies of a program fragment with possible changes.
Type~4 code clones are behavioral copies of a program fragment that do not have any syntactic similarity but implement the same functionality (semantic equivalence).
For example, the iterative and recursive implementations of the Fibonacci algorithm have no syntactic similarity while implementing the same functionality.

Juergens et al.~\cite{Juergens2010} have shown that existing tools only have limited capabilities for detecting Type~4 clones.
This limitation can also be seen in various clone detection tool comparisons \cite{Svajlenko2015, Koschke2007, Roy2007, Bellon2007, Farmahinifarahani2019} through the absence or explicit exclusion of Type~4 clones.
Nevertheless, Type~4 clones exist and tools for detecting them are needed~\cite{Juergens2010, Kafer2018}.

We present \emph{Semantic Clone Detection via Probabilistic Software Modeling (SCD-PSM)}.
SCD-PSM detects semantic clones with no textual and structural similarity.
First, a network of Probabilistic Models (PMs) is built via Probabilistic Software Modeling (PSM)~\cite{Thaller2019a}.
Each PM models an executable (e.g., a method in Java) in the program under analysis.
SCD-PSM leverages these PMs and their inferential capabilities to detect semantically equivalent executables.
Probabilistic inference enables a similarity measure based on probabilities.
These probabilities are used to conduct statistical tests (Generalized Likelihood Ratio Test) that produce the final clone decision.

\section{Background}\label{sec:background}
\begin{figure*}
	\centering
	\begin{subfigure}[b]{0.32\textwidth}
		\includegraphics[width=\textwidth]{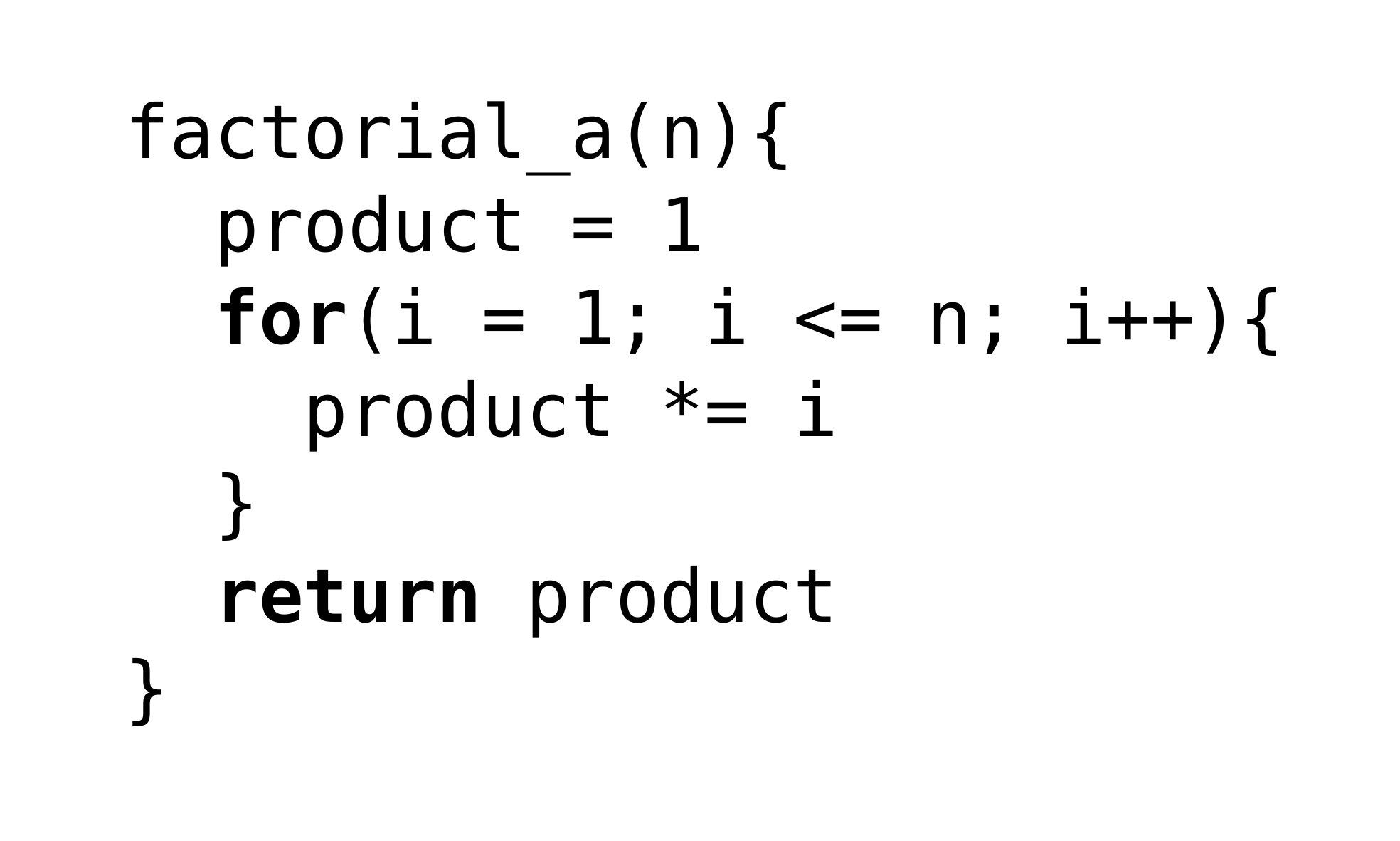}
		\caption{A \emph{for} implementation of factorial.}
		\label{fig:clone 1}
	\end{subfigure} \hfill
	\begin{subfigure}[b]{0.32\textwidth}
		\includegraphics[width=\textwidth]{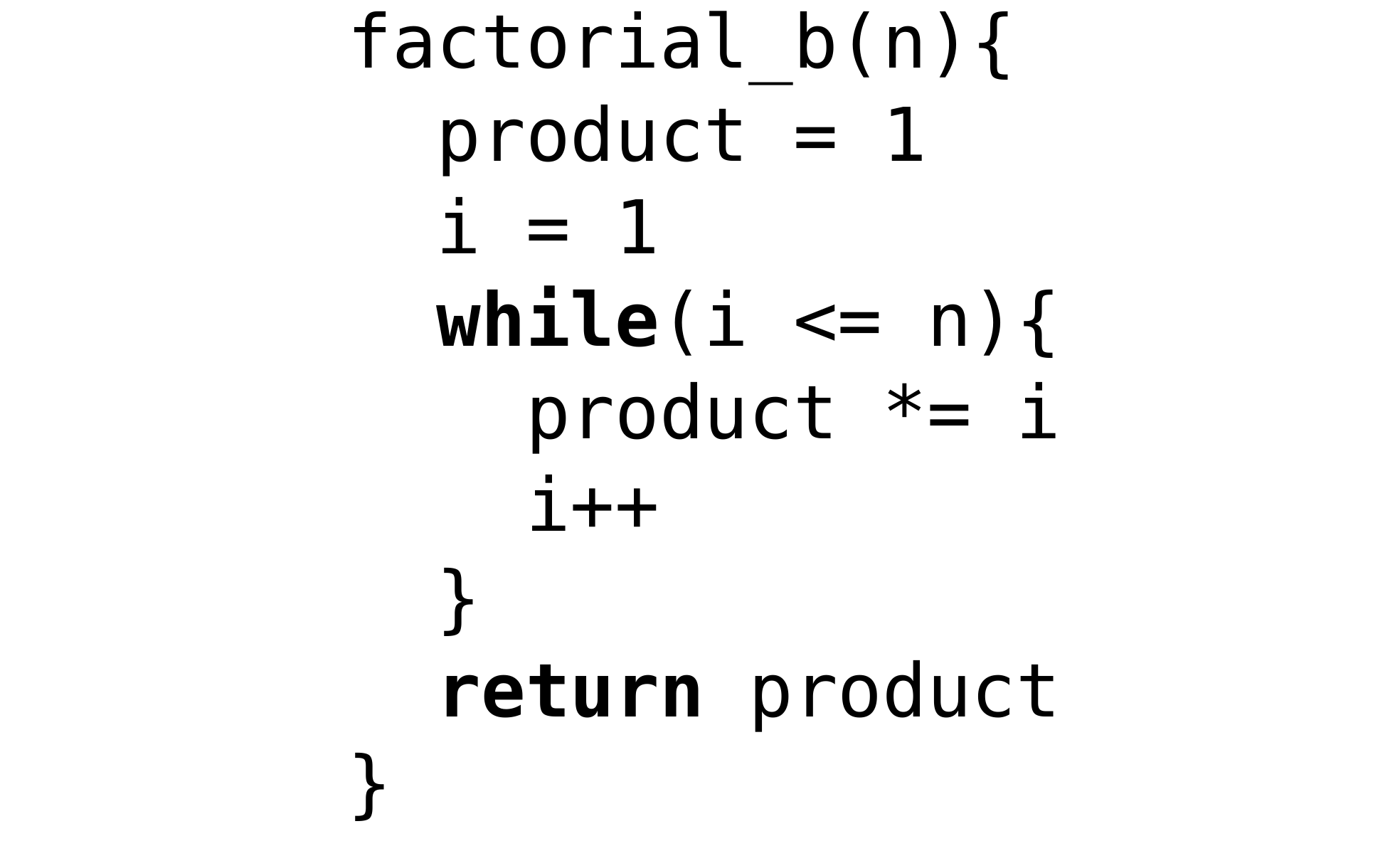}
		\caption{A \emph{while} implementation of factorial.}
		\label{fig:clone 2}
	\end{subfigure} \hfill
	\begin{subfigure}[b]{0.32\textwidth}
		\includegraphics[width=\textwidth]{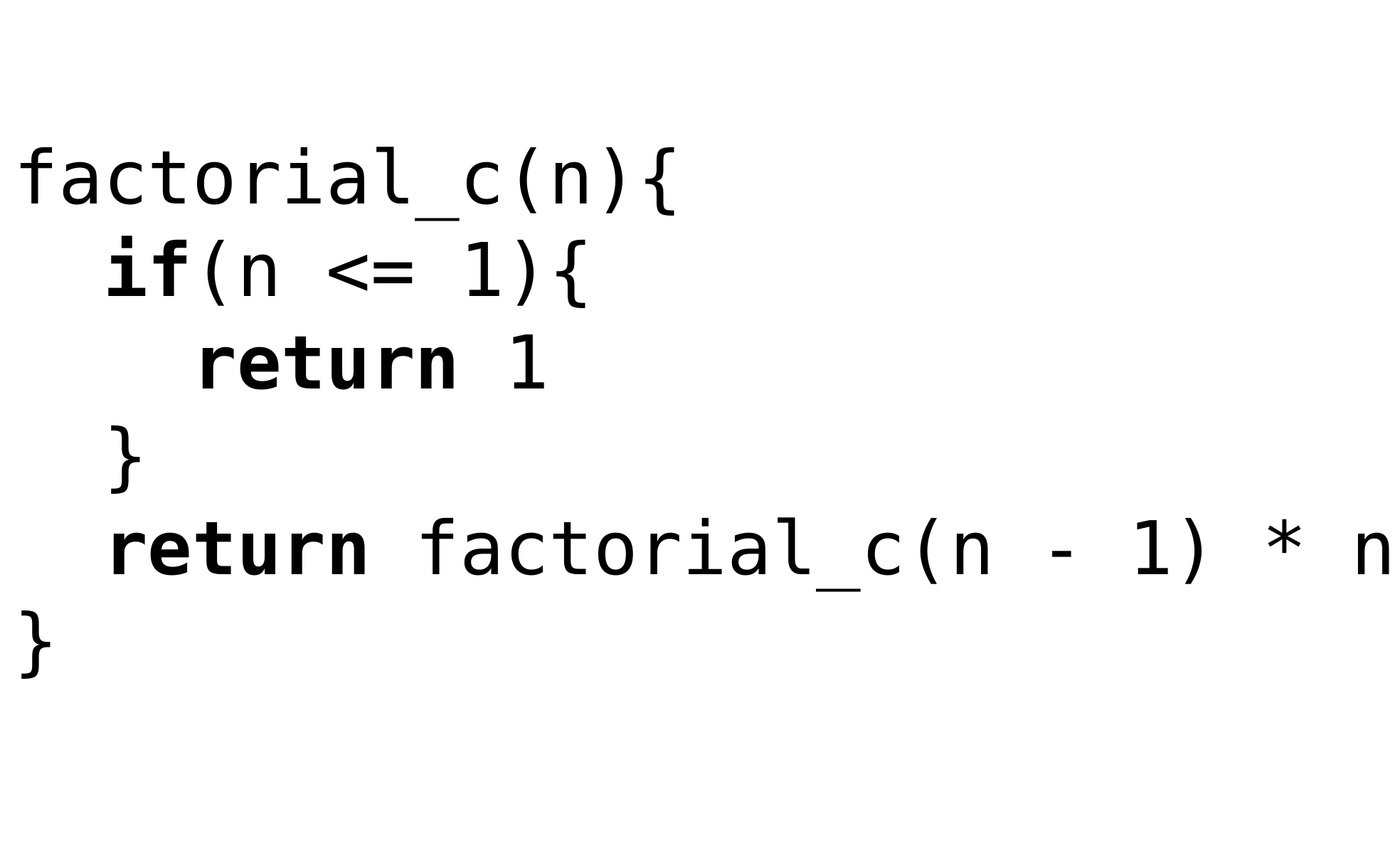}
		\caption{A \emph{recursive} implementation of factorial.}
		\label{fig:clone 3}
	\end{subfigure}
	\caption{
		The \emph{for} and \emph{while} implementations are complex Type~3 clones in which new lines were added and some changed.
		The \emph{recursive} implementation is a Type~4 clone of the \emph{for} and \emph{while} implementations without any syntactic resemblance.
	}
	\label{fig:clone}
\end{figure*}

A basic understanding of clone detection and probabilistic software modeling is needed to understand the approach.
We will use a \code{monospace} font to refer to program elements (e.g., \code{factorial\_a}) and \emph{italics} to refer to the corresponding model elements (e.g., \emph{factorial\_a}).

\subsection{Clone Detection}
Clone detection is the process of finding pairs of similar program fragments as illustrated in Figure~\ref{fig:clone}.
Figure~\ref{fig:clone} shows three different implementations of the factorial computation.
Figure~\ref{fig:clone 1} uses a for loop, while Figure~\ref{fig:clone 2} uses a while loop implementation.
Finally, Figure~\ref{fig:clone 3} uses recursion to compute the factorial of $n$.
The clone detection process includes the representation (e.g., text fragments), pairing (e.g., of text fragments of similar size), the similarity evaluation (e.g., counting the differences in the text fragments), and the clone decision (e.g., less than 10 differences).

\emph{Representations} can be, for example, text (e.g., source code), graphs (e.g., AST), or probabilistic models (like in this work).
\emph{Pairing} is the process of selecting two code fragments that are potentially a clone.
Each pair is called a \emph{candidate clone pair} (or candidate pair).
The \emph{similarity evaluation} measures the similarity between the fragments of a candidate pair.
The \emph{clone decision} labels the candidate pair as a clone given that the similarity fulfills some criteria.

The properties of the similarity metric splits clones into two groups~\cite{Roy2007}.
Type~1-3 clones capture textual similarity while Type~4 clones capture semantic similarity \cite{Bellon2007, Koschke2007, Krinke2001, Roy2007, Rattan2013, Thaller2017}.
These types are increasingly challenging to detect, with Type~4 being the most complex one.
Figure~\ref{fig:clone 1} and Figure~\ref{fig:clone 2} are an instance of a Type~3 clone while  Figure~\ref{fig:clone 1} (or Figure~\ref{fig:clone 2}) and \ref{fig:clone 3} are an instance of a Type~4 clone.
Note, that the definition of a \emph{semantic clone} is often relaxed where up-to $50\%$ syntactic similarity of the code fragments is allowed \cite{Svajlenko2015, Saini2018}.
However, we consider these clones as complex Type~3 clones (additions, deletions, reordering) and \emph{not} as semantic clones.
This means that semantic clones in the context of this work are clones with no syntactic similarity except for per-chance similarities (e.g., equal parameter names).

\subsection{Probabilistic Software Modeling}\label{sec:background psm}
Probabilistic Software Modeling (PSM)~\cite{Thaller2019a} is a data-driven modeling paradigm that transforms a program into a network of Probabilistic Models (PMs).
PSM extracts a program's structure given by types, properties, and executables (e.g., classes, fields, and methods respectively in Java).
This structure includes the call dependencies between the different code elements which defines the topology of the PM network.
Each PM is optimized towards a program execution.
The program execution can either be synthetic (e.g., random testing), from tests (e.g., developer tests), or from the program in its production environment.
In the context of clone detection, synthetic program executions suffice as the results are based on differential comparisons of two elements.

Each PM represents an executable (e.g., a method in Java) in the program.
Inputs are parameters, property reads, and invocation return values.
Outputs are the method return value, property writes, and invocation parameters.
The distinction between inputs and outputs is only a logical view from a software engineering perspective.
The actual PMs are multivariate density estimators without such distinction (joint model of all variables).
PMs can generate observations that are similar to the initial training data.
More importantly, each model can evaluate the likelihood of data.
The likelihood is used to detect behavioral equivalence between models, which is then generalized to the semantic equivalence between executables in the program.

The PMs in the network are real Non-Volume Preserving transformations (NVPs)~\cite{Dinh2016}, a generative likelihood-based latent-variable model for density estimation.
NVPs learn a function that maps data to a known latent-space, e.g., input parameter values \code{n} and return values \code{product} of \code{factorial\_a}, to a bivariate normal distributions.
More formally, each NVP is a neural network that learns a bijective function $f: \bm{X} \mapsto \bm{Z}$ (with $g=f^{-1}$) between the original data $\bm{x} \in \bm{X}$ and predefined latent-variables $\bm{z} \in \bm{Z}$.
The latent-variables are selected, such that sampling, conditioning, and likelihood evaluation is efficient and straightforward, e.g., via an isotropic unit norm Gaussian $\mathcal{N}(0, \bm{1})$.

\emph{Sampling} generates data $\bm{x} \in \bm{X}$ by drawing observations from the latent-variables $\bm{z} \sim \bm{Z}$ and inverting them via the NVP to the original data-space $\bm{x} = g(\bm{z}) \sim \bm{X}$.

\emph{Conditioning} finds a latent-space configuration (i.e., a latent-code) $\bm{\hat{z}}$ such that the associated data  $g(\bm{\hat{z}}) = \bm{\hat{x}}$ satisfies a given condition.
First a proposal code is drawn form the latent-space $\bm{\hat{z}}$ which is then inverted to its data form $\bm{\hat{x}} = g(\bm{\hat{z}})$.
Then the error is measured on the conditioned dimensions via, e.g., Mean Squared Error (MSE).
The error is used to update  the latent code $\bm{\hat{z}}$ and the procedure is repeated until convergence.
For example, one can condition the return value from \code{factorial\_a} on to the return value of the \code{fibonacci} method.
First, samples are drawn from the \emph{factorial\_a} model retaining only the dimension associated with the return value.
Then, samples are drawn from the \emph{fibonacci} model and the error between the return value dimensions is computed. 
This error is back-propagated to the latent-code which is updated according to the errors.
After convergence of the optimization the \emph{fibonacci} sample contains the same return values as imposed by the \emph{factorial\_a} sample.
Furthermore, the remaining dimension $n$ is resampled (imputed) in such a way that it adheres to the joint relationship of all the variables in \emph{fibonacci}.
Finally, \emph{fibonacci} can be used to evaluate the likelihood of the conditioned sample.

\section{Approach}\label{sec:approach}
\begin{figure}
	\centering
	\includegraphics[width=\linewidth]{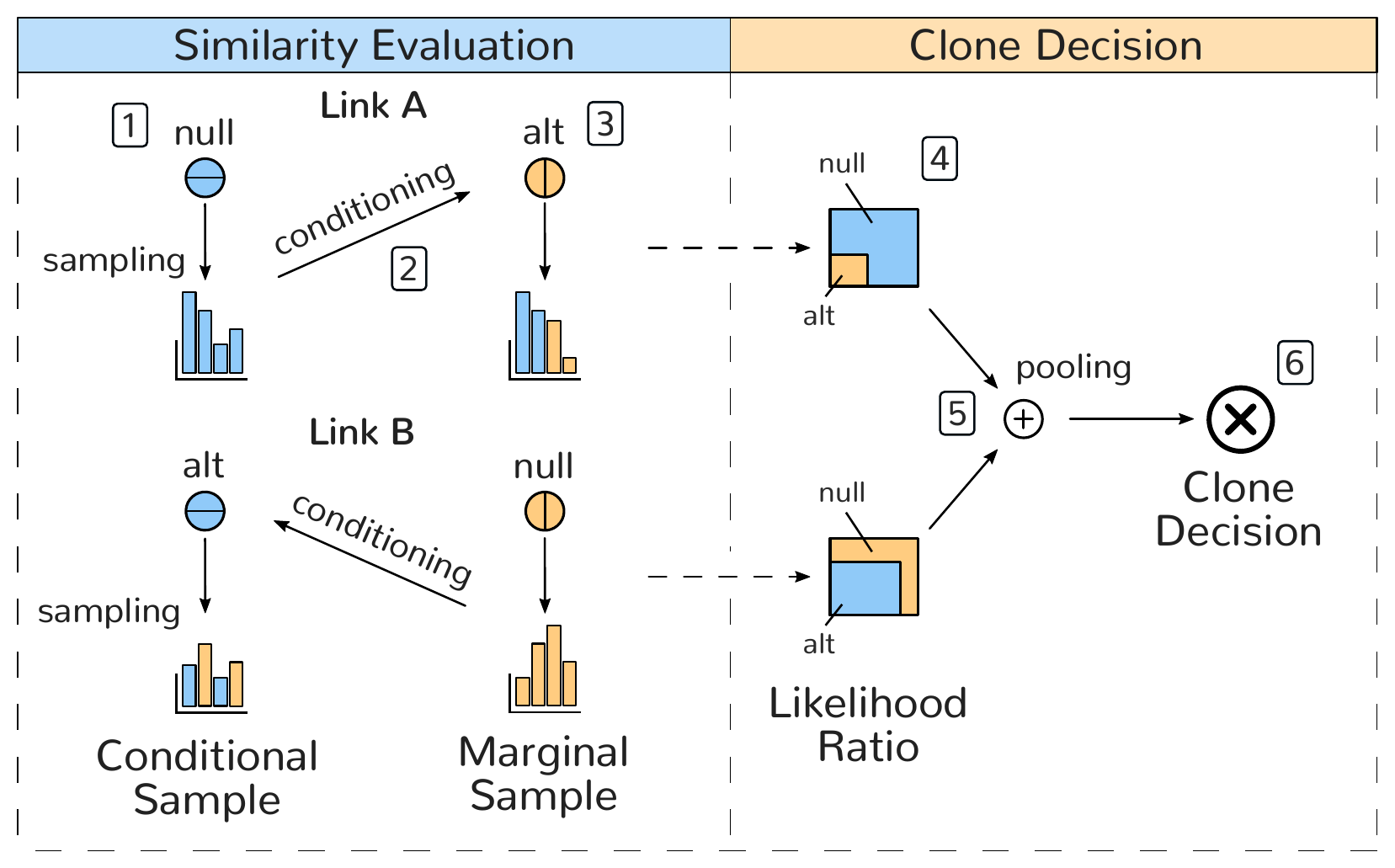}
	\caption{SCD-PSM evaluates the similarity of a pair of models via their data likelihood. The likelihoods are combined into the final clone decision.}
	\label{fig:scd-psm}
\end{figure}

SCD-PSM uses the models built by PSM and compares them for behavioral equivalence.
The behavioral equivalence is then generalized to semantic equivalence of executables (i.e., methods).

\subsection{Similarity Evaluation}\label{sec:similarity}
The similarity evaluation computes the cross-wise likelihood of the models by \emph{sampling} and \emph{conditioning}.
Given is a pair of candidate PMs, each representing an executable.
The similarity evaluation starts by selecting a reference model (null-model) $M^{null}$ and an alternative model (alt-model) $M^{alt}$.
Then, null-dimensions $M^{null}_{\bm{k}}$ and alt-dimensions $M^{alt}_{\bm{k}}$ are selected from the models, e.g., parameter $n$ from \code{factorial\_a} is compared to parameter $n$ of \code{factorial\_b}.
Then, a reference sample $D^{null}_{\bm{k}}$ is generated by $M^{null}$ as illustrated in Figure~\ref{fig:scd-psm} (1) representing the behavior of $M^{null}$.
This reference sample is used to generate a conditioned alternative sample $D^{alt \mid null}$ (2) representing the behavior of $M^{alt}$ given that dimensions $\bm{k}$ are fixed to the behavior of $M^{null}_{\bm{k}}$ (3).
Finally, the likelihood of $D^{null}$ is evaluated under $M^{null}$, resulting in the base likelihood of the reference sample under the null-model $LL^{null}$, and $D^{alt \mid null}$ is evaluated under $M^{alt}$, resulting in the likelihood of the conditioned alternative sample under the alt-model $LL^{alt\mid null}$.
Then, the \emph{null} and \emph{alt} roles are swapped and the procedure is repeated (see Figure \ref{fig:scd-psm} Link B).

The swapping of roles is necessary because of sub-model relationships.
For instance, one model returns data distributed according to $N(0, 1)$ and the other according to $N(0, 5)$.
One link will lead to a high likelihood (sub-model is null) while the other link will result in low likelihood (super-model is null).

In conclusion, the similarity evaluation tests the likelihood of the models in the context of each other.
The final clone decision is based on these likelihood values.

\subsection{Clone Decision}\label{sec:clone analysis}
The final step is to combine the likelihood values from the similarity evaluation to a final decision as shown in Figure~\ref{fig:scd-psm}).
The two \emph{likelihood ratios}  (4) are combined by a \emph{pooling} operator (5) and compared against a critical value yielding the final \emph{clone decision} (6). 

More formally, the procedure makes use of the Generalized Likelihood Ratio Test (GLRT)~\cite{Fan2001}.
The log-GLRT measures whether the log-likelihoods are significantly different from 0 with 
\begin{equation}
	\lambda = LL^{alt} - LL^{null},
\end{equation}
where $LL$ is the log-likelihood.
The null hypothesis is that the models are equal.
It is rejected for small ratios $\lambda \le c$ where $c$ is set to an appropriate Type~1 error, i.e., false-positive rate.
For example, $\lambda < \log(0.01)$ allows 1 out of 100 candidates to be a false positive, i.e., wrongly rejecting semantic equivalence.

The \emph{Clone Decision} (6) is computed by \emph{pooling} (5) the link results.
\emph{Hard pooling} accepts the candidate pair as a clone if the null hypothesis for \emph{both} links could not be rejected.
\emph{Soft pooling} accepts the candidate pair as a clone if the average log-likelihood ratio of both links cannot be rejected.
Hard pooling does not allow any sub-model relationship, while the soft pooling relaxes this condition slightly.

The final requirement is that a candidate pair is only accepted as a clone if the selected dimensions $\bm{k}$ of both, $M^{null}$ and $M^{alt}$ contain at least one input and output dimension.
That is, methods are semantically equivalent if at least parts of their input and output relationship is equivalent.

In conclusion, the clone decision combines the link results and controls the results for a predefined false positive rate.

\section{Study}\label{sec:study}
\begin{table}[!t]
    \centering
    \ra{1.2}
    \caption[Project Overview]{The 8 subject examples used in the evaluation.}
    \label{tab:subjects}
    \begin{threeparttable}
      \resizebox{\columnwidth}{!}{
        \begin{tabular}{@{}l r r r r @{}}
            \toprule
            \textbf{Subject}   &\textbf{Style} &\textbf{Clone Class}  &\textbf{Parameter} &\textbf{Executable} \\
            \midrule    
            Factorial    & iterative   & A & \num{1}  & \num{1}  \\
            Factorial    & recursive   & A & \num{1}  & \num{1}  \\
            Fibonacci    & iterative   & B & \num{1}  & \num{1}  \\
            Fibonacci    & recursive   & B & \num{1}  & \num{1}  \\
            BubbleSort   & iterative   & C & \num{1}  & \num{1}  \\
            BubbleSort   & recursive   & C & \num{3}  & \num{2}  \\
            MergeSort    & iterative   & C & \num{6}  & \num{2}  \\
            MergeSort    & recursive   & C & \num{8}  & \num{3}  \\
            \midrule    
                        &             &  & \num{22} & \num{12} \\
            \midrule
            \bottomrule
        \end{tabular}
        }
    \end{threeparttable}

\end{table}

We implemented a prototype for SCD on top of PSM and applied the similarity evaluation given in Section~\ref{sec:approach}.

\begin{enumerate}
	\item The study uses 8 well-known algorithms listed in Table~\ref{tab:subjects} distributed in 3 clone classes.
    Each clone class is a well-understood example of semantic clones with 0\% syntactic similarity.
    Each subject was triggered with positive uniform distributed random values.
	\item The \emph{Probabilistic Model Network} was computed via Gradient, a PSM prototype~\cite{Thaller2019a}.
	The same hyper-parameters were selected as in our previous reported experiments.
	\item The \emph{Candidate Clone Pairs} were all combinations of dimensions of the PMs.
	The candidate pairs were formed from all 8 subject systems.
	\item Each valid candidate pair was tested for behavioral equality by cross-wise likelihood evaluation described in Section \ref{sec:similarity}.
	\item The clone decision was computed via the GLRT and the results were pooled as described in Section \ref{sec:clone analysis}.
\end{enumerate}

\subsection{Controlled Variables}
The study controls for \emph{pooling}, the \emph{Type~1 error} , and the \emph{number of particles} used in the similarity evaluation (Section~\ref{sec:similarity}).

\begin{description}
    \item [Pooling] describes how likelihoods are combined to the final clone decision \{soft, hard\} (see Section \ref{sec:clone analysis}).
    \item [Type~1] error, or the false-positive rate, defines the critical value $c$ at which clones are considered significantly different \{0.001, 0.01\} (Section \ref{sec:clone analysis}). 
    The critical value is the total Type~1 for both links.
    \item [Number of Particles] are the number of samples that are sampled during the similarity evaluation for the reference sample $D^{null}$ and the alternative sample $D^{alt}$. A low number of particles is faster to compute but has a higher variance in the results.
\end{description}

\subsection{Response Variables}
The performance of the clone detection is measured via \emph{precision}, \emph{recall}, and the \emph{balanced accuracy}.
These metrics are computed by the True Positive (TP), False Positive (FP), True Negative (TN), and False Negative (FN) proportion of detected clone instances, e.g., correctly identifying a clone pair counts towards TP.

\begin{description}
    \item[Precision] measures the performance to detect only relevant instances given by
    \begin{equation}
        \dfrac{TP}{TP + FP}
    \end{equation}
    \item[Recall] measures the performance of detecting all relevant instances given by
    \begin{equation}
        \dfrac{TP}{TP + FN}
    \end{equation}
    \item[Balanced Accuracy] measures the performance of detecting relevant and irrelevant instances but considers a possible imbalance between the number of relevant and irrelevant instances.
    It is given by
    \begin{equation}
        \dfrac{\dfrac{TP}{TP+FN} + \dfrac{TN}{TN + FP}}{2}
    \end{equation}
\end{description}

\subsection{Experiment Results}\label{sec:study results}
\begin{table*}[ht]
    \centering
    \footnotesize
    \ra{1.2}
    \caption[Study Results]{Results of the clone detection experiments.}
    \label{tab:study results}
\begin{threeparttable}
        \begin{tabular}{@{} llrr|rrrrrrr @{}}
          \toprule
         \multicolumn{4}{c}{\textbf{Controlled Variables}} & \multicolumn{7}{c}{\textbf{Response Variables}} \\
         & Pooling & Type I & Particles  & TP & FP & TN & FN & Precision & Recall & Balanced Accuracy \\ 
          \midrule
  1 & hard & 0.001 & 10 &    22 &   0 &  14 &   0 & 1.00 & 1.00 & 1.00 \\ 
  2 & hard & 0.001 & 50 &   18 &   0 &  10 &   8 & 1.00 & 0.69 & 0.78 \\ 
  3 & hard & 0.001 & 100 &   20 &   0 &  12 &   4 & 1.00 & 0.83 & 0.89 \\ 
  4 & soft & 0.001 & 10 &    14 &   0 &  18 &   4 & 1.00 & 0.78 & 0.89 \\ 
  5 & soft & 0.001 & 50 &   22 &   0 &  10 &   4 & 1.00 & 0.85 & 0.89 \\ 
  6 & soft & 0.001 & 100 &   22 &   0 &  10 &   4 & 1.00 & 0.85 & 0.89 \\ 
  7 & hard & 0.010 & 10 &     8 &   0 &  26 &   2 & 1.00 & 0.80 & 0.94 \\ 
  8 & hard & 0.010 & 50 &    14 &   0 &  14 &   8 & 1.00 & 0.64 & 0.78 \\ 
  9 & hard & 0.010 & 100 &    14 &   0 &  14 &   8 & 1.00 & 0.64 & 0.78 \\ 
  10 & soft & 0.010 & 10 &    16 &   2 &  10 &   8 & 0.89 & 0.67 & 0.72 \\ 
  11 & soft & 0.010 & 50 &   20 &   0 &  12 &   4 & 1.00 & 0.83 & 0.89 \\ 
  12 & soft & 0.010 & 100 &    22 &   0 &  14 &   0 & 1.00 & 1.00 & 1.00 \\ 
           \midrule
           \bottomrule
        \end{tabular}
\end{threeparttable}

\end{table*}

The study results are given in Table~\ref{tab:study results}.
Average precision was \num{0.991}, recall was \num{0.797}, and the balanced accuracy was \num{0.870}.
The precision across experiments was excellent, indicating that models can reliably detect behavioral equality.
This is reflected in the low number of FPs.
However, the FNs indicate that some positive examples are missed.
Reducing the Type~1 error, i.e., falsely rejecting semantic equality, improves on the FNs.
The recall was good for most evaluates.
However, hard pooling caused a slight drop in the recall.
The balanced accuracy is good to excellent for most experiment configurations.
Perfect scores are given for experiment 1 and 12.

No effect of Pooling, Type~1, and Particles on the accuracy can be seen.

\section{Discussion}\label{sec:discussion}
The results from Section~\ref{sec:study results} are encouraging.
The general performance was good to excellent.
No significant difference between the different levels of pooling, Type~1, and Particles in Table \ref{tab:study results} can be seen.
However, a larger sample size is needed to precisely attribute effects on the performance.
In the $10$-particle setting a higher variance of performance can be seen caused by per-chance errors.
The number of FPs is in all experiments low which is expected given that the Type~1 error was set to $0.001$ and $0.01$.
In contrast, the number of FNs is acceptable.
This is reflected in the Recall that ranges from $0.64$ to $1$.
The balanced accuracy shows high detection rates of the approach in most experiments settings.

\section{Limitations}\label{sec:limitations}
SCD-PSM inherits the limitations of PSM.
PSM models data.
Object references are handles to containers (objects) that store data.
Thereby, SCD-PSM cannot detect semantic clones of executables that solely manage object references, e.g., a collection library.
However, this limitation does only hold if the program never accesses the underlying data.
Furthermore, PSM explodes lists into singular values since distributions do not contain any order information.
This means executables that change the order of sequences are matched based on the values, not their order.
As a consequence, invoking a wrongly implemented, e.g., sorting algorithm, would result in a false positive.
Extending PSM to model distributions of sequences will alleviate this issue.

A limitation of the detection process is that it is built on runtime observations.
This means that the approach can only be applied to runnable source code.

The final limitation is that the approach cannot detect Type 2-3 clones.
Slight changes, e.g., flipping a plus sign to a minus, have large implications on the resulting runtime behavior.
These changes will impact the semantic detection process such that the candidate clone pair will not be accepted.
For example, common clone detectors will report Listing \ref{lst:inc} and Listing \ref{lst:dec} as clones since they differ only by one character (ignoring names and reducing minimum size).
However, this does not hold for Type~4 detectors because the input and output relationship is different.
In contrast, many clone detectors will not detect Listing \ref{lst:inc} and \ref{lst:inc1} as clones because of the many additions.
Type~4 detectors will report this pair as clones since the behavior of adding one to the input is identical.
This hints that Type 2-3 and Type 4 clones represent detached concepts that share less common ground than expected.
More importantly, this raises the question whether existing detectors that report Type~3-4 detection capabilities generalize as expected.

\begin{lstlisting}[caption={Increment method}, label={lst:inc}]
inc(a: Int): Int{
    return a + 1
}
\end{lstlisting}

\begin{lstlisting}[caption={Decrement method}, label={lst:dec}]
dec(a: Int): Int{
    return a - 1
}
\end{lstlisting}

\begin{lstlisting}[caption={Complicated increment method}, label={lst:inc1}]
inc(a: Int): Int{
    b = 1 * 3.12
    c = b / 2
    d = c + -0.5
    return (Int) a + d
}
\end{lstlisting}

\section{Related Work}\label{sec:related work}
Many studies have evaluated textual clones.
However, there are only a few studies reporting reliable results on semantic clones without relaxing the definition of Type~4.

Rattan \cite{Rattan2013} et al. provided a review of clone detection studies.
The review also investigated approaches that tackle Type~4 clones.
They conclude that some approaches solve approximations (i.e., complex Type~3 clones) of Type~4 clones.

Horwitz \cite{Horwitz1990} detected textual and semantic differences in programs via a Program Representation Graph, which is similar to a Program Dependency Graph (PDG).
PDG-based approaches \cite{Krinke2001, Gabel2008, Komondoor2001} use (static) data and control dependencies to find similar sub-graphs between the candidates.
They can detect complex Type~3 clones, e.g., Figure \ref{fig:clone 1} and Figure \ref{fig:clone 2}.
However, the compared PDG sub-graphs are a representation of the source code; thereby, the approaches still rely on syntactic similarity \cite{Wagner2016}.

Another category of semantic clone detectors are test-based methods.
Test-based methods randomly trigger the execution of two candidates and measure whether equal inputs cause similar outputs.
Jiang and Su \cite{Jiang2009} were able to detect semantic clones without syntactical similarities.
A similar approach was presented by Deissenboeck et al. \cite{Deissenboeck2012}.
One issue with test-based clone detection is that candidates need a similar signature.
Differences in data types or the number of parameters can not be effectively handled by the test-case generators or the similarity measurement.
SCD-PSM works similar to test-based methods in that it observes the runtime and compares the resulting behavior.
However, SCD-PSM builds generative models from the observed behavior capable of generating and evaluating data.
Missing dimensions are imputed by conditioning and sampling.
This allows SCD-PSM to overcome the issue of signature mismatches.
Furthermore, PSM abstracts the data types into text, integer, and floats mitigating data type mismatches.

Finally, the clone detector Oreo \cite{Saini2018} has also reported Type~3 to Type~4 detection capabilities.
Oreo uses a combination of representations and detection stages to find clones.
Most important is the semantic similarity comparison based on \emph{actions} a method takes, e.g., accessing an array, writing a property, or invoking a method.
These actions correspond, to some extend, to the dimensions of PSM models, i.e., represent entry-points of information (e.g., field accesses, invocations, etc.)
Oreo counts these entry-points and compares then between the fragments in a candidate pair.
No analysis of the runtime assignments is conducted, nor is the relationship between the actions analyzed like SCD-PSM does.
Oreo reports many complex Type~3 and Type~4 clones up to 50\% syntactic similarity based on this semantic similarity (and the additional pipeline steps).
However, more research is needed to identify the weaknesses and strengths of both approaches.
This highlights the need for a hard but well understood baseline dataset of Type~4 clones similar to the examples in our study but extended with a larger variety of semantic clones.

\section{Conclusion and Future Work}\label{sec:conclusions}
In this work, we presented a viable approach for semantic clone detection - Semantic Clone Detection via Probabilistic Software Modeling (SCD-PSM).
SCD-PSM leverages the PMs of PSM to detect method level semantic clones with 0\% syntactic similarity.

We have discussed the similarity evaluation and the clone decision that represent the central aspect of a clone detector.
We evaluated the concepts on a set of well-known semantic clones that provide a hard baseline for Type~4 detectors.

Our future work is to evaluate the scalability of the approach with large programs.
Furthermore, we want to compare SCD-PSM with existing Type~3 clone detectors.

In conclusion, SCD-PSM is capable of detecting semantic clones with 0\% syntactic similarity.

\section*{Acknowledgments}
\ifunblind
The research reported in this paper has been supported by the Austrian ministries BMVIT and BMDW, and the Province of Upper Austria in terms of the COMET - Competence Centers for Excellent Technologies Programme managed by FFG.
\else
The research reported in this paper has been supported by <blinded ORG>. 
\fi


\balance

\bibliographystyle{IEEEtran}
\bibliography{references}

\end{document}